\documentclass[doublecol]{epl2} 
\title{Pentagon deposits unpack under gentle tapping}
\shorttitle{Pentagon deposits} %Insert here a short version of the title if it exceeds 70 characters

\author{Ana M. Vidales\inst{1,2} \and Luis A. Pugnaloni\inst{3} \and Irene Ippolito\inst{4}}
\shortauthor{Ana M. Vidales \etal}

\institute{
  \inst{1} Laboratorio de Ciencia de Superficies y Medios Porosos, Departamento de F\'{\i}sica, Universidad Nacional de San Luis y CONICET\\
  \inst{2} Groupe Mati\`ere Condens\'ee et Mat\'eriaux, UMR CNRS 6626, Universit\'e de Rennes I, F-35042 Rennes Cedex, France\\
  \inst{3} Instituto de F\'{\i}sica de L\'{\i}quidos y Sistemas Biol\'{o}gicos (UNLP-CONICET), cc. 565, 1900 La Plata, Argentina\\
  \inst{4} Grupo de Medios Porosos, Facultad de Ingenier\'{\i}a, Universidad de Buenos Aires,
Paseo Col\'{o}n 850, 1063 Buenos Aires, Argentina and CONICET (Argentina).
}
\pacs{81.05.Rm}{Porous materials; granular materials}

\pacs{45.70.Cc}{Granular flow: mixing, segregation and stratification}

\pacs{87.18.Bb}{Computer simulation}

\abstract{
We present results from simulations of regular pentagons arranged in a rectangular die. The particles are subjected to vertical tapping. We study the
 behavior of the packing fraction, number of contacts and arch distributions as a function of the tapping amplitude. Pentagons show peculiar features
 as compared with disks. As a general rule, pentagons tend to form less arches than disks. Nevertheless, as the tapping amplitude is decreased, the typical
 size of the pentagon arches grows significantly. As a consequence, a pentagon packing reduces its packing fraction when tapped gently in contrast
 with the behavior found in rounded particle deposits.
}

\begin{document}

\maketitle

\section{Introduction}
The study of compaction of granular matter under vertical tapping is a subject of much debate and consideration. Setting apart all issues related to the
 slow relaxation shown by these systems, and considering only the steady state regime whenever achieved, the investigated granular deposits attain
 rather high packing fractions when tapped very gently. Moreover, these systems display a rapid reduction in the packing fraction as tapping intensity is
 increased followed by a smooth increase at large intensities. On the one hand, The initial decrease in packing fraction has been observed in simulation of spheres \cite{Barker1}, in
3D experiments with glass beads \cite{Nowak1,Richard1} and in simulation of disks \cite{Pugnaloni1,Pugnaloni4}. On the other hand, the smooth increase at
 large intensities has been pointed out in simulations of disks \cite{Pugnaloni1,Pugnaloni4} and in experiments on 2D packings \cite{Blumenfeld1}. 
 Also, a hint of this smooth increase can be appreciated in early 3D experiments (see figure 2 in Ref. \cite{Nowak1}). In some cases, the steady state may be
 obtained by extended constant intensity tapping; however, other experimental configurations may require a suitable annealing in order to achieve the
 so called 'reversible branch' \cite{Richard1}.

Up to now, few studies of this type have been carried out on pointed objects. A first experimental investigation uses assemblies of spheres
to build up more complex objects which however retain the smooth edges of the constituents \cite{Rankenburg1}.

Here, we show that packings of pentagons simulated through a pseudo molecular dynamics method (PMDM) present a response to vertical tapping
 which is significantly different from that observed in packings of rounded grains like disks or spheres. We base our assertions on the comparisons with
 disk packings obtained with an analogous method.

It is worth mentioning that studies on pentagon assemblies do exist \cite{Limon1,Limon2,Schilling1}. These make special emphasis on the crystallization of
 these systems. However, these experiments and simulations consider systems which relax continuously under the effect of a background vibration (either thermal
 or mechanical agitation).

\section{The model}
The main satages of our simulations consist in: (a) the generation of an irregular base, (b) sequential deposition of pentagons to create an initial
 packing, (c) vertical tapping obtained through vertical expansion followed by small random rearrangements, and (d) non-sequential (simultaneous) deposition of the 
pentagons using a PMDM.

We sample $1000$ regular pentagons from a uniform size distribution ($5\%$ dispersion). A number of them are placed at the bottom of a rectangular
 die in a disorder way in order to create an irregular base. Arranged in this manner, the $N$ base particles fix the wall-to-wall width of the die which is
 about 40 particle diameters. These
 pentagons remain still over the course of the tapping protocol. The remaining pentagons are poured one at a time from the top of the die and from random
 horizontal positions and random orientations. Each grain falls following a steepest descendent algorithm.
 When a pentagon touches an already deposited particle, it is allowed to rotate about the contact point until a new contact is made or until the contact
 point no longer constrains the downward motion of the particle which is deemed to fall freely again. If a particle has reached two contacts such that the 
$x$-coordinate of its center of mass lies between them, the pentagon is considered stable. Otherwise, the pentagon will be allowed to rotate around the
 contact point with lower y-coordinate. Side walls are considered without friction.

Once the initial configuration is obtained, a tapping process is carried out by using an algorithm that mimics the effect of a vertical tap of amplitude
 $A$. The system is expanded by vertically scaling all the $y$-coordinates of the particle centers by a factor $A>1$. Base particles are not subjected
 to this expansion. We then introduce a horizontal random noise for those particles touching
any of the walls of the die. This is done by attempting to displace each of these particles a random distance in the range $[0,A-1]$ towards the center
 of the die in the $x$-direction. Only if the new position of a pentagon does not originate an overlap with neighbor pentagons the move is accepted. Each
 of these particles has only one chance to move. This process mimics in some way the shaking that grains suffer
 in a real experiment because of the collisions with the walls. Note that the amplitude of the random moves is proportional to the amplitude of the
 expansion.

\begin{figure}
\onefigure[width=3.5in]{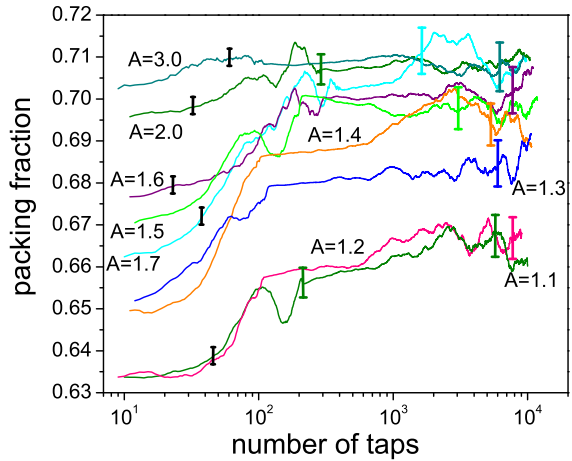}
\caption{Packing fraction of pentagons as a function of the number of taps. The different curves correspond to different amplitudes $A$, as indicated on the plot.}
\label{fig1}
\end{figure}

After expansion and random rearrangements, the particles are allowed to deposit non-sequentially (i.e. simultaneously rather than one at a time) following an algorithm
similar to that designed by Manna and Khakhar for disks \cite{Manna1,Manna2}. In brief, this is a pseudo dynamic method that consists in small falls and rolls
of the grains until they come to rest by contacting other particles or the system boundaries. Particles are moved one at a time but they perform only small moves that do
 not perturb to a significant extent the ulterior motion of the other particles in the system. For very small particle displacements this method yields a realistic 
simultaneous deposition of the grains. Details on the convergence of the results for decreasing values of the size of the particle displacements will be presented elsewere.

Once all pentagons come to rest, the system is vertically expanded again and a new cycle begins. After a large number of taps, the packing attains a steady
 state whose characteristic parameters fluctuate around equilibrium values.

We study the packing fraction $\phi$, coordination number $<z>$, and the arch size distribution $n(s)$ of the deposits. To identify arches one needs
 first to identify the two supporting particles of each pentagon in the packing. Then, arches can be identified in the usual
 way \cite{Pugnaloni1}: we first find all \textit{mutually stable particles} ---which we define as directly connected--- and then
 we find the arches as chains of connected particles. Two pentagons A and B are mutually stable if A supports B and B supports A. Unlike disk deposits
 generated through PMDM, pentagon packings present capriciously shaped arches.

\section{Results and discussion}
The packing fraction of the pentagon deposits is plotted against the number of taps for various tapping amplitudes in Fig.~\ref{fig1}. It is clearly seen that, as the amplitude is
increased, compaction is enhanced. This trend is similar to the one found by Knight et al. \cite{Knight1}, where they observed an increase in the packing fraction with
 the tapping intensity. However, our system reaches a clear plateau after a moderate number of taps irrespective of the tapping amplitude while
in Ref. \cite{Knight1} the steady state was hardly achieved for high tapping amplitudes and definitely not reached for low $A$. Experiments in 2D packings
 \cite{Lumay1} of disks show a much faster equilibration than the 3D packings of Ref. \cite{Knight1}. 

\begin{figure}
\onefigure[width=3.5in]{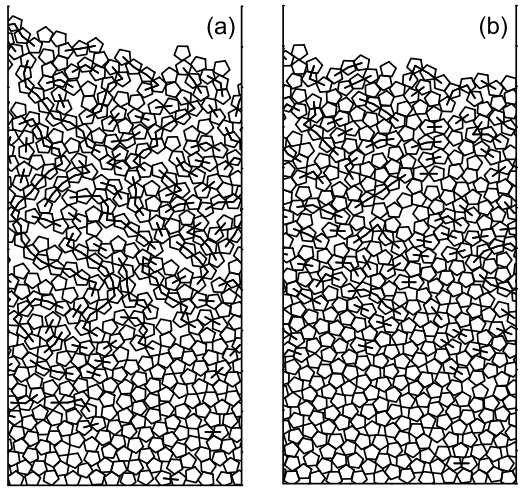}
\caption{Examples of two packing tapped during $5x10^{3}$ times. We only show part of the 1,000 particles assemblies. Arches are indicated by
 segments. (a) $A=1.2$ (b) $A=1.7$.}
\label{fig2}
\end{figure}

We show snapshots of part of two packings in Fig.~\ref{fig2}. Part $(a)$ shows a picture
of part of the whole assembly of a deposit of 1000 pentagons after being shaken $5\times 10^{3}$ times at $A=1.2$. Part $(b)$ shows the same situation but
for $A=1.7$. Arches formed among particles are indicated by segments and will be discussed below.  It can be seen that the final equilibrium
positions of the particles in each case are quite different. At low $A$, the creation of long arches
due to blocked rollings of the particles gives as a result a lower $\phi$ in comparison with that shown for a packing tapped at higher amplitudes. Moving
 the particles farther appart during expansion allows them to rearrange better and to increase side-to-side contacts.

\begin{figure}
\onefigure[width=3.5in]{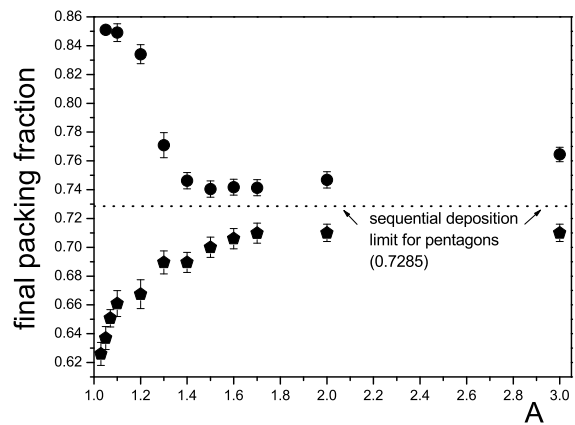}
\caption{Final values of the packing fraction obtained by averaging over the last $1,000$ taps as a function of the tapping amplitude for disks (circles)
 and pentagons (pentagons). The horizontal dotted line corresponds to the sequential deposition limiting case for pentagons (see text for details).}
\label{fig3}
\end{figure}

In Fig.~\ref{fig3} we plot the final values of $\phi$, obtained when the system attains the steady state regime (averaging over the last $1000$ taps)
 as a function of the tapping amplitude. We compare results with the same experiment carried out on disks \cite{Pugnaloni1} (using the same size
 dispersion, number of particles and die size) and with a pentagon limiting
 case obtained as follows. We rise all pentagons up to a large height and let them fall again, one at a time and in order of height (the lower particle
 first). This process leads to the highest compaction. The tapped deposits approach this value of $\phi$ when $A$ is increased, as seen in
 Fig.~\ref{fig3}. Since the deposition is sequential, pentagons do not form arches at all in the limiting case.

There are two clear distinctions between the behavior shown by disks and that displayed by pentagons. Firstly, disks attain larger packing fractions at all
 tapping amplitudes. This is to be expected since pentagons, if not carefully arranged, tend to leave large interstitial spaces. Secondly, while disks
 present a nonmonotonic dependence of $\phi$ versus $A$, pentagons show a monotonic increase in the packing fraction. At high values of $A$ both
 systems increase $\phi$ with increasing tapping amplitudes and eventually reach a maximum plateau value. For low $A$ we find that disks tend to order
 and so increase $\phi$ as $A$ is decreased \cite{Pugnaloni1}. A minimun in the packing fraction of disks is then located at intermediate values of $A$. However, this
 feature is not present in pentagon packings. Pentagons seem not to order at low $A$, and $\phi$ does not present a minimum as in disk packings.

Realistic molecular dynamic simulations of the tapping of pentagon packings yield higher densities overall due to the particular thermal like vibrations this
 type of simulation suffer until equilibrium is achieved \cite{Patrick}. This feature resembles the high densification of packings obtained in Ref.
 \cite{Limon1}. The same effect is observed when realistic molecular dynamics of disks \cite{Pugnaloni4} are compared with corresponding
 PMDM \cite{Pugnaloni1}. However, PMDM has been shown to yield the same general trends observed in realistic molecular dynamics (compare Ref. \cite{Pugnaloni4}
 with Ref. \cite{Pugnaloni1}).

\begin{figure}
\onefigure[width=3.5in]{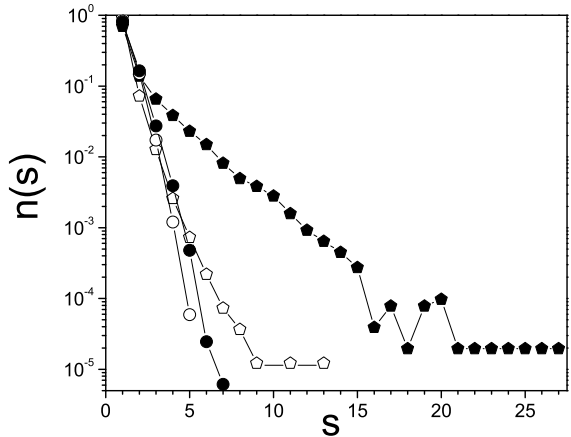}
\caption{Distribution of arch sizes for disks (circles) and pentagons (pentagons) at $A=1.2$ (filled symbols) and $A=3.0$ (open symbols).}
\label{fig4}
\end{figure}

In pentagon packings, we find that the number of arches presents a monotonic decrease with increasing $A$ in contrast with the behavior of disks that
 present a maximum at the same tapping amplitude where the minimum packing fraction is achieved. It is particularly interesting that at $A < 1.1$
 disks enter an ordered phase \cite{Pugnaloni1} where arches are largely eliminated from the system whereas the pentagon deposits remain in a disordered
 state with an increasing number of arches up to very small tapping amplitudes.

We observe that, in general, pentagons form less arches than disks (about $40\%$ less arches). This seems to be contradictory with the fact that we found that
pentagons show a lower coordination number. However, this effect is explained by the wider arch size distribution found in pentagons. In Fig. \ref{fig4}
 we show the distribution of arch sizes for pentagons and disks at two values of $A$. We confirm here that for low $A$ pentagons
have a larger tendency to form large arches (up to 20 particles), whereas disks form arches of less than 10 particles. A detailed study of the
 particle-particle contacts and the formation of arches will be presented elsewhere.

In order to assess whether the tapping protocol applied to the packings is significant in the results discused above, we have carried out an annealed tapping
 on our packings to compare with the constant tapping discused up to this point. We start from a sequentially deposited packing and then tap the system
 at variable amplitude. The amplitude was increased from $A=1.1$ to $A=1.7$ in steps of $0.1$ and $5,000$ taps where applied at each amplitude value. Then,
 the same protocol was followed but for decreasing amplitudes. No evidence of hysteresis nor irreversibility is found in our results. We also found that
the annealing curves  coincide with the constant tapping
 results of Fig. \ref{fig3}. Both, disks and pentagons, attain a unique packing fraction value for given tapping amplitude no matters the history of the
 tapping protocol.

\begin{figure}
\onefigure[width=3.5in]{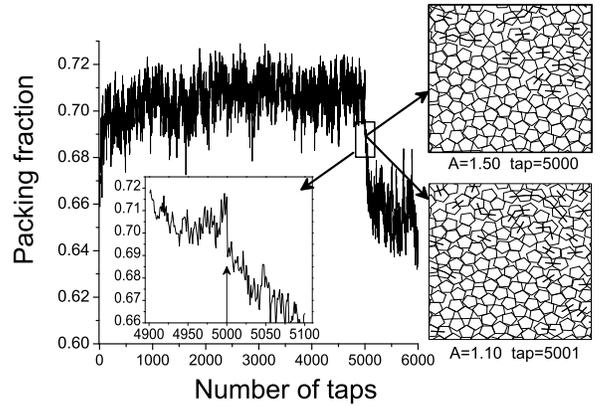}
\caption{Packing fraction of pentagons as a function of the number of taps before and after a sudden reduction in tapping amplitude. From $t=1$ to
 $t=5000$ the packing is tapped with $A=1.5$, from $t=5000$ on the amplitude is set to $A=1.1$. The insets show snapshots of parts of the system
 (with arches indicated by segments) before and after the change in tapping amplitude and a magnification of the main graph.}
\label{fig4-5}
\end{figure}

Previous simulations on disks \cite{Pugnaloni1} and experiments on glass beads \cite{Nowak1} do show an irreversible branch in this type of experiment. It
 is important to note that in the case of previous simulations \cite{Pugnaloni1} the annealing was conducted in a different manner since the tapping amplitude was
 increased in a quasi-continuum fashion and a single tap was applied at each value of $A$. This prevented the disk packing from reaching the steady state
 at each value of $A$. In the present work we give sufficient time for the system to reach the steady state at each amplitude. On the other hand, the
 annealing experiments by Nowak et al. \cite{Nowak1} were conducted in much the same way as our simulations, however, their system presented a very slow
 relaxation that effectively prevented the packing from 'equilibration' at each tapping amplitude.

To get a closer insight into the 'peculiar' behavior of pentagons (i.e. the reduction of packing fraction as $A$ diminishes)
 we show in Fig. \ref{fig4-5} the evolution of a pentagon deposit after a sudden reduction in tapping amplitude. After $5000$ taps applied to the
 system with $A=1.5$ we set the tapping amplitude to $A=1.1$ and continue to tap the deposit for $1000$ taps. As we can
 observe, the reduction in $A$ induces a rapid reduction in packing fraction associated with an increase in the size of
 the arches formed (see insets in Fig \ref{fig4-5}), in contrast with the behavior generally observed in deposits of disks.
 This seemingly paradoxical effect is in fact simple to explain. Arches ---which are the main void-forming structures--- are more easily
 created when particles start deposition from an initial high density expanded configuration. At low $A$, the expanded configuration leaves particles very
 close to each other and this make particles to meet each other more often during deposition, enhancing the probability of arch formation. This has been
 discused recently by Roussel et al. \cite{Roussel1} and it has been observed by Blumenfeld et al. \cite{Blumenfeld1} in experiments of compaction in two
 dimensional granular systems.

\section{Conclusions}

\begin{figure}
\onefigure[width=3.5in]{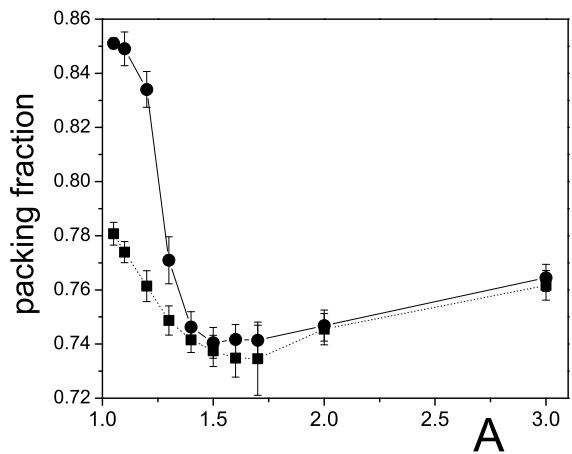}
\caption{Packing fraction for disks as a function of $A$ for two values of size dispersion: $5 \%$ (circles), and $50 \%$ (squares).}
\label{fig5}
\end{figure}

We have shown that, for pentagons, either through constant tapping or annealing, the steady state of the packing presents a monotonically increasing
 packing fraction with tapping intensity. However, disks and spheres display a clear initial reduction in the packing fraction as tapping intensity is
 increased followed by a smooth increase at large amplitudes \cite{Barker1,Nowak1,Richard1,Pugnaloni1,Pugnaloni4,Blumenfeld1}.
 Such finding reveals that the complexity of pentagon deposition leads to an unexpectedly simpler behavior of the packing fraction as compared with simpler
 systems.

To our understanding, the behavior of rounded particles ---which increase density on reduction of tapping intensity--- are indeed puzzling; while pentagons
 seem to behave as expected. If grains fall from a highly compact expanded configuration they should form more arches, and hence reduce packing fraction. 
 Rounded particles do not follow this pattern as has been observed in experiments and simulations of various kinds. Although the behavior of rounded particles
 seems to be considered as reasonable for most workers, no thorough discusion on this has been given in the literature. Most authors explain the effect on the basis that
 large taps create voids but do not explain how these voids are created from a mechanical point of view. According to the detailed discussion presented by Roussel et
 al. \cite{Roussel1} large taps should destroy arches (and then voids). We believe that the 'reasonable' behavior is that large taps eliminate arches and voids;
 however, at low tapping amplitudes, we presume that this phenomenon competes with the crystal-like ordering that reduces arch formation in disk packings in our
 simulations.

We have tested the hypothesis that partial
 ordering leads the nonmonotonic behavior of disks and spheres. However, some trial simulations carried out with rather polydisperse disks that are known
 to show frustration of order still present the same nonmonotonic features, although less marked than in monosized disks (see Fig. \ref{fig5}). A sensible
 explanation for the formation of large arches at low tapping
 amplitude should in principle shed light on this issue. At present we can only suggest that pentagons (and any other pointed particles) have a larger
 tendency to multiparticle collisions. Multiparticle collisions are necessary (although not sufficient) to form many-particle arches. These multiparticle
 collisions are enhanced by two factors: (a) the fact that pentagons may approach each other closer than disks (recall that a side-to-side contact
 leaves pentagon centers separated by $\approx 0.8$ particle diameters) which increases number density despite the lower packing fraction, and (b) the
 associated collisions on rolling originated by the protruding vertices. A recent model based on collisional probabilities \cite{Roussel1} for the formation
 of arches may help to quantify these effects.

\acknowledgments
AMV thanks to the Groupe Mati\`ere Condens\'ee et Mat\'eriaux and to the Universit\'e de Rennes I, France, for their hospitality and support
during this investigation. Special thanks to Luc Oger, Patrick Richard and Rodolfo U\~nac for valuable suggestions and discussions. LAP acknowledges
 financial support from CONICET (Argentina). AMV and II acknowledge financial support from CONICET (Argentina) under project PIP 5496.

\end{document}